# Método para medição de uniformidade de engajamento de alunos: medindo antes e depois da introdução de exercícios com planilhas gráficas

**George C. Cardoso**[1]

[1] Universidade de São Paulo/FFCLRP/Departamento de Física, gcc@usp.br

**Resumo**

*O uso de planilhas eletrônicas no ensino e aprendizado de ciências e matemática tem sido feito em várias ocasiões e com resultados positivos. Entretanto, estudos quantitativos sobre quais alunos se beneficiam com as planilhas eletrônicas são raros ou inexistentes. Neste trabalho mostramos um estudo que compara o aprendizado de uma turma de estudantes usando método tradicional com turma similar que usou planilhas gráficas em algumas aulas, laboratórios e exercícios. Utilizamos atividades de laboratório computacional com planilhas eletrônicas para introduzir conceitos de cálculo e física para alunos de Biologia antes de introduzir tais conceitos de maneira formal. Em ambos semestres, cada aluno foi avaliado no início por um teste sobre conceitos básicos necessários para o acompanhamento das aulas. Essa avaliação visou mensurar o nível de cada aluno e a estatística de heterogeneidade da turma. As avaliações iniciais também foram usadas para comparação e cálculos de correlação com os desempenhos finais do semestre. Nossos resultados mostram que com o uso das planilhas gráficas não há mudança relevante na correlação entre o desempenho no teste no início do semestre com o desempenho final na disciplina, quando a turma inteira é considerada. Entretanto, para a metade da turma inicialmente abaixo da mediana há melhora significativa na correlação entre o nível de conhecimento inicial e o desempenho semestral na disciplina. Observações sugerem que esse aumento da correlação é devido a um aumento no engajamento dos alunos inicialmente abaixo da mediana, em comparação com o engajamento de tais alunos num semestre convencional. Concluímos que no ensino de física e matemática, a incorporação de exercícios de visualização gráfica e numérica é importante principalmente para engajar alunos com deficiências matemáticas iniciais. Entretanto, não houve diferença quantitativa para aqueles alunos inicialmente acima da mediana nos grupos estudados.*

**Palavras-chave**: Planilhas eletrônicas; Excel; Avaliação de metodologia de aprendizado; engajamento de alunos

## Introdução

No ensino de ciências exatas para alunos ingressantes em cursos de graduação, conceitos simples podem ser de difícil compreensão. Para alguns alunos a linguagem matemática da física e do cálculo pode ser comparada a uma língua estrangeira desconhecida para eles. Um exemplo simples onde pode ocorrer dificuldade é a identificação dos pontos críticos de uma função usando o cálculo. A simples nomenclatura, tais como pontos de inflexão, primeira derivada e segunda derivada, pode sobrecarregar alguns alunos fazendo-os perder o foco da lógica matemática a ser aprendida.

Neste trabalho discutimos as diferenças encontradas no aprendizado de turmas antes e depois da introdução da planilha eletrônicas como ferramenta auxiliar no ensino de física e matemática. O *software* Microsoft Excel foi escolhido por ser uma ferramenta disponível nos computadores da universidade. Planilhas eletrônicas gratuitas equivalentes tais como a do LibreOffice foram também utilizadas por vários estudantes. A literatura especializada mostra que já existiram outros esforços na direção do uso do planilhas gráficas para facilitar o aprendizado dos conceitos relevantes de matemática (STIELER, 2007; DE NOVAES, 2009; RIBEIRO, 2016), de física (REDISH, 2000; SUBRAMANIYAN, 2013) e de matemática aplicada à biologia (DONOVAN, 2002). Os resultados citados pela literatura focam em mostrar o aumento no engajamento dos estudantes e/ou no aprendizado dos conceitos estudados através do uso de planilhas eletrônicas.

A proposta inovadora deste trabalho é avaliar quantitativamente o aumento no engajamento e desempenho dos estudantes em uma disciplina de física e matemática com o uso de planilhas eletrônicas. Dois semestres em anos subsequentes são avaliados, sendo o primeiro deles sem uso do Excel. A referência usada na comparação quantitativa entre os dois semestres são notas dos alunos em avaliações de 16 questões sobre matemática básica, aplicada no início de cada semestre.

Não existe consenso na literatura sobre a definição do termo engajamento em ensino e aprendizagem, havendo várias formas de defini-lo tais como engajamento intelectual, emocional e comportamental (PUGH, 2010; TROWLER, 2010). Tais definições são baseadas em questionários qualitativos ou semi-quantitativos (KUH, 2008). Para os nossos fins, definimos engajamento como a correlação entre o potencial inicial do aluno com o aprendizado e o desempenho ao longo da disciplina. Engajamento assim definido é sinônimo de um esforço genuíno do aluno de forma que, em média, cada aluno atinge seu potencial relativo revelado em um teste inicial.

Os dados foram obtidos a partir das notas em provas, trabalhos e listas de exercício. Apesar da observação do comportamento da turma mostrar um melhor engajamento qualitativo com uso de Excel, a análise quantitativa da turma como um todo não mostrou diferenças significativas para com o método convencional. Uma reanálise dos dados mostrou que no semestre convencional não houve nenhuma correlação das notas semestrais da metade da turma com notas iniciais abaixo da mediana com seu nível de conhecimento no teste inicial. Com o uso do Excel, a metade da turma com nota no teste inicial abaixo da mediana passou a ter suas notas semestrais melhor correlacionadas com seu grau de conhecimento inicial. Para a metade da turma com grau de conhecimento no teste inicial acima da mediana, o grau de correlação do conhecimento inicial com desempenho final foi o mesmo independentemente do tipo de semestre (convencional ou com uso de Excel). Com base nesses resultados, levantamos a hipótese de que o uso de planilhas faz com que mesmo alunos com mais deficiência mantenham esforço e interesse para atingir seus potenciais relacionado com o grau habilidade ou conhecimento mostrado no teste inicial, enquanto o método tradicional faz muitos dos alunos abaixo da mediana desistirem de se esforçar por encontrarem dificuldade. Isso se reflete na falta de correlação do nível de conhecimento anterior com o desempenho final, que propomos como métrica para avaliar engajamento.

**Metodologia**

O estudo foi conduzido com turmas da disciplina Fundamentos de Física e Matemática para Biologia na FFCLRP (Universidade de São Paulo, campus de

3Ribeirão Preto), cuja carga horária é de 4 horas semanais. Os alunos são majoritariamente do primeiro ano do curso.

Em 2015, a disciplina foi ministrada com metodologia convencional para ensino de cálculo e de física, usando demonstrações e exercícios a serem resolvidos sem uso de computador. No semestre com uso de planilha eletrônica (2016) essas atividades foram introduzidas com uso de Excel, assim como vários dos exercícios para casa faziam uso das planilhas. Para cada tema principal da disciplina, antes de se introduzir métodos analíticos convencionais, utilizamos práticas de laboratório com o Excel. Houve um total de cinco aulas em laboratório de computação. Essas práticas aconteceram na semana anterior às aulas teóricas onde o tópico era desenvolvido com mais rigor.

Durante o semestre, as primeiras aulas de laboratório foram as mais importantes em relação ao uso do programa propriamente dito, pois o Excel era novidade para a maioria dos alunos. A seguintes atividades de laboratório de computação foram executadas durante o semestre:

1) Introdução ao uso de planilhas: construção de gráficos e cálculo de limites;
2) Cálculo das retas tangentes de gráficos de funções;
3) Cálculo de áreas sob curvas;
4) Modelo epidemiológico SIR;
5) Ajuste de funções: função logística e predição de populações.

Nos laboratórios, cada tópico foi apresentado de forma intuitiva usando a matemática mais simples possível e o Excel como ferramenta de visualização. O problema a ser resolvido era explicado e um exemplo de uso era dado; ao final, os alunos tinham cerca de 45 minutos para praticar exemplos sob orientação do docente e de um monitor. Uma lista de problemas em Excel para cerca de uma hora de trabalho era dada para estudo extraclasse. Solicitava-se um relatório de pesquisa bibliográfica sobre tópicos do tema (por exemplo: "Integrais de Riemann"). Pedia-se que tal relatório utilizasse pelo menos seis dos exemplos/gráficos trabalhados durante o laboratório ou no exercício para casa. O uso do próprio trabalho para ilustrar o relatório minimizava a probabilidade de plágio e valorizava o esforço do estudante. O relatório ajudava os alunos a se prepararem para a aula teórica com os conceitos básicos e a linguagem formal do tema. Isso caminha no sentido das tendências da *flipped classroom* ou "aula invertida", um método que incentiva os alunos a estudarem o tópico com antecedência para participarem mais ativamente durante as aulas (ABEYSEKERA, 2015).

O uso do computador na disciplina possibilitou aplicações interessantes que normalmente seriam muito complexas para o cálculo analítico, aumentando a percepção de relevância dos temas abordados para problemas em Biologia. A dificuldade em encontrar exercícios que lembrem aplicações práticas em cursos básicos de física e matemática para Biologia é um problema recorrente quando usamos os métodos e livros convencionais. Isso se dá porque os problemas que ilustram aplicações interessantes são complexos e demorados para serem resolvidos sem auxílio de um computador.

Os testes adotados no início dos semestres para documentar o estado inicial de cada estudante duravam 15 minutos e consistiam de questões abertas cujas soluções eram óbvias, triviais e imediatas para aqueles que tinham conhecimento dos conceitos avaliados. Os testes foram realizados sem aviso prévio. As perguntas versavam sobre conceitos simples que se esperava que o aluno soubesse no início do semestre, incluindo, por exemplo, soma de duas frações, potenciação, logaritmos, funções trigonométricas e esboço de gráficos simples. Tais testes eram aplicados, recolhidos dos alunos e redistribuídos para que colegas pudessem corrigir para retorno imediato das respostas corretas. Após correção, eram feitos histogramas no quadro com o número de acertos por questão e foram discutidos em detalhe as questões mais problemáticas da turma. Toda a atividade durava cerca de 40 minutos. Como identificação, somente os números de matricula eram listados no teste.

Durante o semestre foram realizadas duas provas convencionais, sem consulta. Atividades de listas de exercícios ou relatórios eram passadas semanalmente. Em sete das semanas do semestre houve pelo menos duas questões na lista de exercícios a serem resolvidas com uso de Excel. Em ambos os semestres estudados havia atividades de monitoria e abertura do docente para consulta com os alunos.

**Resultados**

Na Figura 1 temos a distribuição das notas do teste inicial nos semestres convencional (2015) e com Excel (2016). O gráfico está construído de forma que no eixo horizontal as notas estão em ordem crescente e no eixo vertical temos a distribuição de probabilidade cumulativa das notas dos estudantes. A comparação entre as curvas é similar à usada no teste de Kolmogorov-Smirnov. A informação importante que pode ser observada da Figura 1 é que não há diferença significativa entre as distribuições de nota ou nível de conhecimento prévio (condições iniciais) das turmas de 2015 e de 2016. Isso faz com que as comparações dos resultados dos dois métodos de aprendizado sejam relevantes. Noutras palavras, as distribuições estatísticas das notas das duas turmas são idênticas. Isso nos causa pouca surpresa porque apesar das turmas serem diferentes – um único aluno de 2015 estava presente na turma de 2016 – o número de alunos que fez o teste foi acima de 60 e cada turma vem da mesma população de alunos. Em 2015 a média do teste inicial foi 6,1 com desvio padrão 2,4 (n = 61) e em 2016 a média do teste inicial foi 5,8 com desvio padrão 2,3 (n=64).

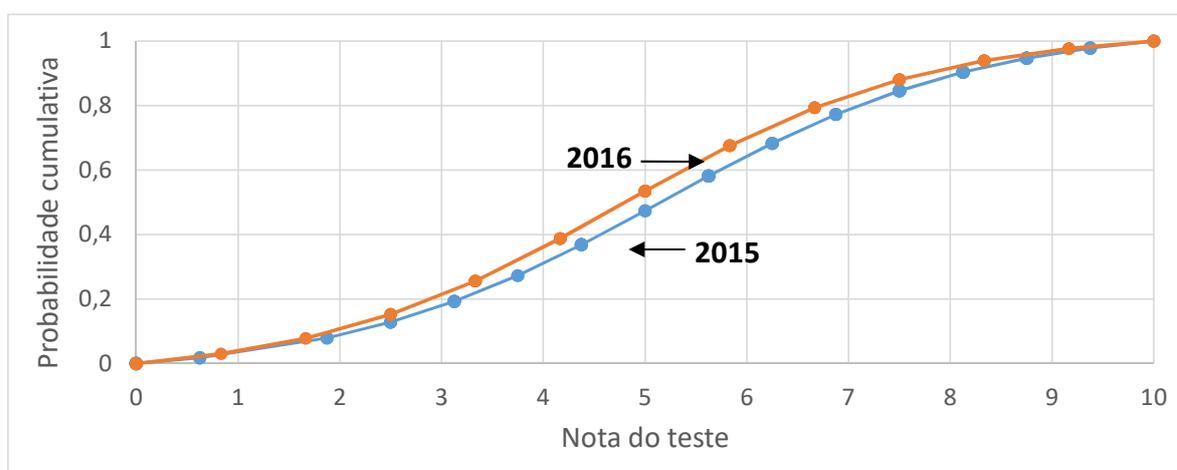

**Figura 1:** *Probabilidades cumulativas vs. notas dos testes sobre conhecimentos básicos realizados no início dos semestres letivos de 2015 (curva de baixo) e 2016 (curva de cima).*

Na Figura 2 temos as médias de final de semestre das turmas convencional (2015) e usando Excel (2016), ambas antes da prova de recuperação. Observa-se que existe uma melhoria de aproximadamente um ponto na média da turma quando as planilhas foram empregadas. Como as avaliações durante o semestre foram diferentes nos anos de 2015 e 2016, não podemos concluir nada com essa diferença.

Na Figura 3, estudamos as correlações das notas dos testes iniciais com o desempenho no final do semestre. Observamos que no semestre com uso de Excel existe uma correlação maior entre conhecimento inicial e desempenho final. Entretanto, o coeficiente de correlação de Pearson é raiz de $R^2$, e como mostrado na Tabela 1, a diferença de correlação para as turmas não é significante pois apenas o primeiro digito da correlação é significante para amostras da ordem de 10 a 100.

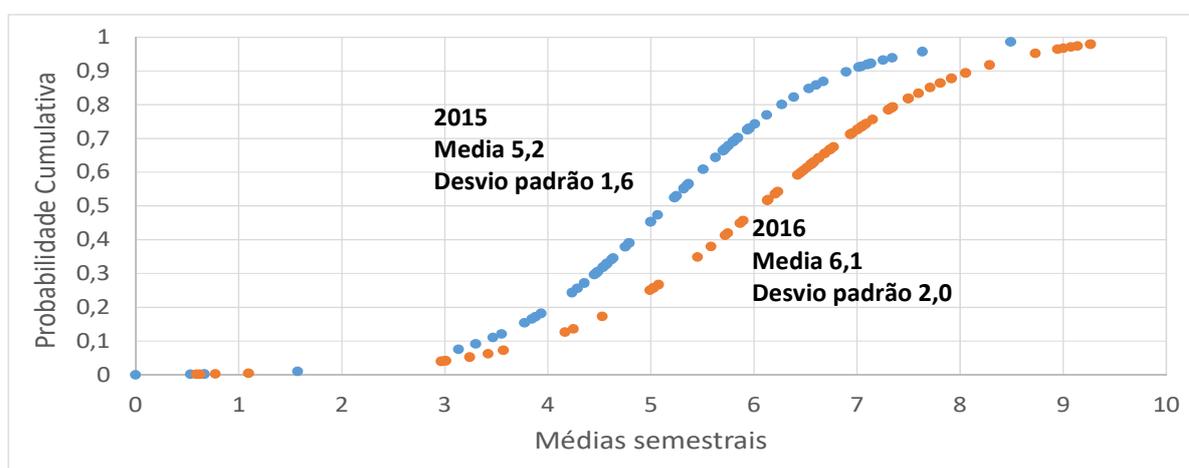

**Figura 2:** *Probabilidades cumulativas vs. médias semestrais, antes da prova de recuperação (REC) para 2015 (curva de cima) e 2016 (curva de baixo).*

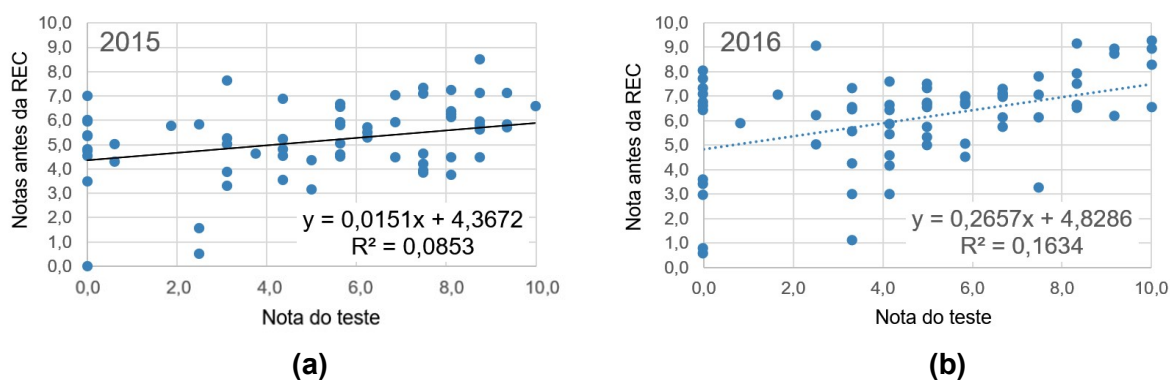

(a)  (b)

**Figura 3:** *Correlação das notas dos testes iniciais (Nota do teste) com notas dos alunos antes da prova de recuperação (REC). (a) Resultado de semestre convencional e (b) Resultado de semestre usando gráficos em planilha eletrônica.*

66Na Tabela 1 mostramos as correlações em mais detalhes dividindo cada uma das turmas em duas metades: a metade com notas dos testes iniciais acima da mediana, e a metade abaixo da mediana. As correlações não chamam a atenção exceto pelo fato de que no semestre convencional (2015) a metade da turma com nível de conhecimento no teste inicial abaixo da mediana tiveram notas totalmente descorrelacionadas com o nível de conhecimento prévio. A correlação das notas da metade melhor preparada da turma em 2015, embora baixa, foi quatro vezes maior que a correlação da metade que ficou abaixo da mediana em conhecimento prévio. No semestre onde métodos gráficos com planilhas de Excel foram adotados (2016) a correlação do desempenho de ambas as metades da turma com suas condições iniciais foi a mesma. Comportamento similar foi encontrado entre os grupos de alunos que foram aprovados direto e alunos que terminaram o semestre com média abaixo de 5,0 (ficaram em recuperação).

**Tabela 1:** Coeficientes de correlação de Pearson da nota semestral com condição inicial

|  | **2015 (convencional)** | **2016 (Excel)** |
|---|---|---|
| Turma inteira | 0,29 | 0,40 |
| Aprovados direto | 0,32 | 0,28 |
| Em recuperação | 0,15 | 0,30 |
| 50% inicialmente < mediana | 0,05 | 0,44 |
| 50% inicialmente ≥ mediana | 0,22 | 0,40 |

**Discussão**

Com o uso de métodos gráficos e numéricos usando planilhas eletrônicas, o curso fica bem mais interessante e o aprendizado é mais efetivo. A metade da turma inicialmente abaixo da mediana foi capaz de obter desempenho compatível com seus níveis relativos de aptidão ou conhecimento inicial encontrados no início do curso. A falta de tal correlação poderia ser entendida como desinteresse pela disciplina ou falta de esforço genuíno compatível com capacidade. Isso é o que foi verificado nos resultados de 2015, usando métodos tradicionais. Para 2015, a tabela 1 mostra que há perda da correlação entre o nível inicial de aptidão e o desempenho no curso para a metade da turma inicialmente abaixo da mediana. Em ambos os casos (2015 e 2016), as metades das turmas inicialmente acima das medianas tiveram seus desempenhos muito mais correlacionados com o nível de conhecimento inicial. É interessante observar que em 2016 a correlação da turma com o nível de conhecimento inicial foi bem maior, indicando, possivelmente, um maior engajamento dos estudantes com o método de ensino que usa planilhas eletrônicas e métodos gráficos e numéricos.

Em geral um curso que só inclui conceitos simples solúveis com lápis e papel pode levar a acreditar que a matemática e a física estão muito distantes da Biologia, onde os exemplos são relativamente complexos quando colocados em forma matemática. O uso de Excel, possibilitou que os alunos trabalhassem problemas tais como modelos com equações diferenciais acopladas do tipo Lotka-Volterra (dinâmica de populações) e modelos epidemiológicos para planejamento de vacinação de populações.

Em questionários distribuídos ao final do semestre, os alunos indicaram que com o uso de planilhas eles sentiram que tiveram um bom aprendizado da parte analítica da disciplina, e de como aplicar o cálculo e a física em alguns problemas reais. Outro ganho aparente é a noção de que os alunos aprenderam uma das principais ferramentas de trabalho do cidadão contemporâneo que é a planilha eletrônica. A planilha eletrônica gráfica também se mostrou uma forma lúdica de verificar soluções através dos gráficos que eram modificados com simples mudanças de parâmetros, aumentando o interesse e satisfação dos alunos mesmo nas atividades mais complexas.

Como mostrado na seção anterior, o uso da correlação do potencial inicial do aluno com o resultado final parece útil para verificar o engajamento na disciplina. Esse cálculo de engajamento permite também verificar se tanto os alunos mais avançados quanto os mais deficientes no início do curso mantiveram um nível adequado e similar de engajamento. Os níveis de correlação encontrados neste estudo são baixos porque existem vários fatores externos que influenciam no engajamento dos estudantes tais como as demandas das outras disciplinas, diferenças de interesses individuais, fatores emocionais e mudanças no estilo de vida ao entrar na universidade.

O coeficiente de correlação de Pearson R varia entre -1 e +1. Um valor positivo indica que as notas normalizadas do grupo de alunos melhoraram em relação às do teste inicial. Isso é esperado porque o teste inicial foi feito sem aviso prévio enquanto para as avaliações durante o semestre os alunos puderam se preparar. Um coeficiente de correlação alto não significa que, por exemplo, os alunos inicialmente abaixo da mediana terão um grau de aprendizado deficiente; simplesmente significa que a probabilidade do um dado aluno inicialmente abaixo da mediana manterem o rank relativo em relação a seus colegas do mesmo grupo é alta. Entretanto, a média do grupo de alunos que estava inicialmente abaixo da mediana pode ser alta. Um coeficiente negativo indicaria que os melhores alunos no teste inicial tenderiam a tornarem-se relativamente piores no desempenho do semestre, o que seria contra-intuitivo.

Uma das limitações do método de estimativa do grau de engajamento através da determinação da correlação é a necessidade d o teste inicial de avaliação ser adequado para medir o nível de conhecimento e treinamento inicial para a medição do potencial do estudante na disciplina em questão.

**Conclusões**

Usando tarefas com planilhas eletrônicas houve melhor aprendizado e maior engajamento dos alunos, quando comparados com uma turma em um semestre convencional. Em particular, o que encontramos neste trabalho é que houve uma melhora significativa na correlação do nível de conhecimento inicial da metade da turma inicialmente abaixo da mediana com seu desempenho no final do semestre. Acreditamos que esse resultado esteja associado a possibilidade de melhor engajamento dos alunos inicialmente abaixo da mediana, comparado com um semestre convencional. Isso parece acontecer porque a planilha eletrônica permite a realização das operações matemáticas de forma automática e também a visualização imediata de resultados de forma gráfica, facilitando o aprendizado. A possibilidade de alterar parâmetros de um problema e observar mudanças imediatas no resultado final gera um conteúdo intuitivo e mesmo lúdico que mantém o interesse dos alunos.

___